\documentclass{article}

\usepackage{microtype}
\usepackage{graphicx}
\usepackage{subfig}
\usepackage{booktabs} 
\usepackage{hyperref}
\usepackage[font=small,skip=0pt]{caption}

\usepackage[accepted]{icml2019}

\icmltitlerunning{ How to Evaluate Trading Strategies: Single Agent Market Replay or Multiple Agent Interactive Simulation? }

\begin{document}

\twocolumn[
\icmltitle{How to Evaluate Trading Strategies: Single Agent Market Replay or Multiple Agent Interactive Simulation?}

\begin{icmlauthorlist}
    \icmlauthor{Tucker Hybinette Balch}{jpmny}
    \icmlauthor{Mahmoud Mahfouz}{jpmldn,imperial}
    \icmlauthor{Joshua Lockhart}{jpmldn}
    \icmlauthor{Maria Hybinette}{uga}
    \icmlauthor{David Byrd}{gtc}
\end{icmlauthorlist}

~\\

\icmlaffiliation{jpmny}{J. P. Morgan Artificial Intelligence Research, New York, USA}
\icmlaffiliation{jpmldn}{J. P. Morgan Artificial Intelligence Research, London, United Kingdom}
\icmlaffiliation{imperial}{Department of Electrical and Electronic Engineering, Imperial College London}
\icmlaffiliation{uga}{Department of Computer Science, University of Georgia, Athens, USA}
\icmlaffiliation{gtc}{School of Interactive Computing, College of Computing, Georgia Institute of Technology}
\icmlcorrespondingauthor{Tucker Balch}{tucker.balch@jpmorgan.com}
\icmlkeywords{multi-agent systems,  limit order book, market microstructure}
]

\printAffiliationsAndNotice{} 


\begin{abstract}
We show how a multi-agent simulator can support two important but distinct methods for assessing a trading strategy: Market Replay and Interactive Agent-Based Simulation (IABS).  Our solution is important because each method offers strengths and weaknesses that expose or conceal flaws in the subject strategy.  A key weakness of Market Replay is that the simulated market does not substantially adapt to or respond to the presence of the experimental strategy. \emph{IABS} methods provide an artificial market for the experimental strategy using a population of background trading agents.  Because the background agents attend to market conditions and current price as part of their strategy, the overall market is responsive to the presence of the experimental strategy. Even so, \emph{IABS} methods have their own weaknesses, primarily that it is unclear if the market environment they provide is realistic.  We describe our approach in detail, and illustrate its use in an example application: The evaluation of market impact for various size orders.
\end{abstract}

\section{Background and Related Work}

Most professional investors, hedge funds, investment institutions and banks prefer to test trading strategies in simulation before ``going live'' with funds at risk.  A key reason of course is to gain assurance that the strategy is likely to be profitable, or at least that it will not lose money.  It is worth pointing out that not all trading strategies are aimed at producing profits from price moves, some strategies are aimed at minimizing the costs of a transaction.  For instance, a pension fund's management may have concluded that it should reduce its holdings in a particular stock and therefore trigger a sell order for that asset. If this order were sent to an exchange as a market sell order, the price would likely fall significantly and provide the seller a lower average price than they would hope. So brokerages and banks create execution (trading) strategies to minimize that impact, such as discriminating a larger order across as set of smaller orders over time. In general we would like to provide a robust means for evaluating trading strategies.  This motivates our research.

{\bf Research Question:} How can we leverage historical data and Interactive Simulation to most effectively assess experimental trading strategies?

In this paper we will examine two approaches to market simulation that can be used to evaluate an experimental strategy: Market Replay and Interactive Simulation. \emph{Market Replay} is by far the most prevalent method. With Market Replay, historical data is revealed to the experimental strategy as simulated time advances.  Market Replay in the literature is often referred to as to this as \emph{backtesting.} When the strategy chooses to buy or sell at a particular time the backtester executes the order at the current price, where the meaning of ``current price'' varies with the sophistication of the backtester. We will describe a few such backtesting (or Market Replay) techniques in section 2.2. 

From an evaluation point of view, Market Replay has the problem that the simulated market is not responsive to the experimental trading strategy.   For example, in a learning trading strategy trained in Market Replay may learn to exploit specific price histories or conditions that would not exist if the market did adapt or respond.  It is possible that the performance of a strategy refined in Market Replay may be optimistic, and that it would not perform as well in a ``responsive'' real market.

With Interactive Agent-based Simulation (IABS), an artificial market environment is created by populating the simulation with dozens or perhaps thousands of ``background agents.''  Each background agent follows its own private, perhaps randomized, strategy for placing buy and sell orders.  The experimental strategy can then experience and trade in this environment.  Potential advantages of the IABS approach include: that participating market agents will react to the experimental strategy with different consequential orders; that the experimental strategy can be exposed to conditions and situations that may not have occurred historically; and that a much larger corpus of market data can be engaged for strategy evaluation.  Thus IABS may provide a more comprehensive evaluation than would be possible using only historical data.


\subsection{Limit Order Books}

\begin{figure}
 \centering
 \includegraphics[width=0.45\textwidth]{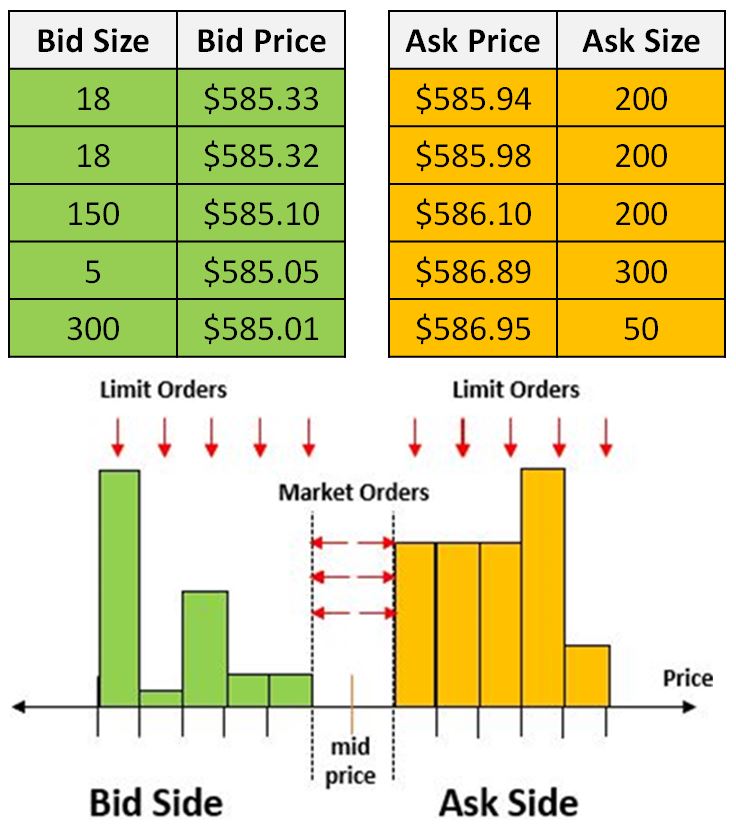}
 \caption{An example limit order book.}
 \label{fig:LOB}
\end{figure}

We briefly review how public exchanges such as NASDAQ and the New York Stock Exchange operate because they are essential to our work.  An exchange facilitates the buying and selling of assets by accepting and satisfying buy and sell orders.   Order types are further distinguished between \emph{limit} orders and \emph{market} orders. A limit order includes a price that should not be exceeded in the case of a buy, or should not be gone below in the case of a sell.  A market order indicates that the trader is willing to accept the best price available immediately. Figure \ref{fig:LOB} shows the difference between limit and market orders in terms of their interaction with the order book (an exchange has an order book for each asset traded). 

The limit order book (LOB) represents a snapshot of the supply and demand for an exchange traded instrument at a given time (see figure \ref{fig:LOB}). It is an electronic record of all the outstanding buy and sell limit orders organized by price levels. The LOB is split into two sides; the ask and bid sides containing all the sell and buy limit orders respectively. The difference between the lowest ask price (best ask) and highest bid price (best bid) is called ``the spread.'' The mid price is the average of the best bid and ask prices. Finally, ``liquidity'' refers to the ease of buying or selling without a significant impact on the price of the instrument. Very liquid instruments tend to have a tight spread with orders arriving in sub microsecond intervals. 

A matching engine is used to match incoming buy and sell orders. This typically follows the price/time priority rule \cite{Nasdaq}, whereby orders are ranked first according to their price. Multiple orders having the same price are then ranked according to the time they were entered. If the price and time are the same for the incoming orders, then the larger order gets executed first.  The matching engine uses the LOB to store pending orders that could not be executed upon arrival. It is important to note that exchanges allow for cancellations or partial cancellations of unmatched limit orders which can lead to complex order book dynamics that arise due to the frequency of these order cancellations. We highlight this as part of the exploratory data analysis section of this paper.

The dynamical properties of limit order books have enjoyed significant coverage in the economics and statistical literature. The survey \cite{gould2013limit} provides a detailed overview of order book models ranging from studies of auction dynamics resulting from interaction of informed and uninformed traders \cite{kyle1985continuous} to mathematical models utilizing Poisson \cite{cont2010stochastic} and Hawkes processes \cite{toke2011market}. \cite{lehalle2018marketmicrostructure} provides an in-depth overview of market microstructure, focusing on topics such as market design, order book dynamics, market impact and the consequences of recent regulatory requirements on market microstructure. 
The topic of market impact and optimal liquidation strategies has been studied extensively in the literature and are covered in \cite{bouchaud2010price} and \cite{cont2014price}. \cite{grinold1995active} covers the sigma-root-liquidity model (square root law) which takes into account the spread cost, daily volatility, daily volume and number of traded shares to estimate market impact. However, this only considers the size of the trade in relation to the daily volume and does not consider other aspects such as how the trade is executed and the rate at which the trades are placed. A number of models exist in the literature which follow the dynamics of the model proposed by \cite{grinold1995active}. These include the continuous time propagator model explained in \cite{gatheral2010no} and the Alfonsi and Schied order book model by \cite{alfonsi2010optimal}. In the continuous time propagator model, the evolution of the stock price is modelled using two functions with one describing the instantaneous market impact and the second being a decay kernel. In \cite{alfonsi2010optimal}, the authors show that an optimal liquidation strategy is bucket-shaped and would involve placing two block of orders at the beginning and end of the proposed trade duration with a constant rate of order placement in-between.

\subsection{Approaches to Backtesting}

Backtesting a trading strategy  allows for evaluating the performance of a strategy in a simulated environment using historical data. Backtesters have different levels of sophistication. A simple form of a backtester would involve the use of the mid price or the last traded price as the historical price at which an evaluated strategy would execute. This approach ignores the fact that in reality, the price at which a trade is executed is not the mid price or historical price but the prices available at the best bid or ask depending on the direction of the trade. Another issue centers on the assumption that the executed trade would not impact the evolution of the historical prices. Therefore, a more sophisticated backtesting approach would implement some form of a market impact model that would take into consideration the size of the trade among other variables to account for their effect on the historical prices. In our experiments, we implement a backtester using a market replay mechanism that does present the experimental strategy with some market impact as a consequence of sizeable orders, thus partially overcoming the issues described with static historical data.

\subsection{Interactive Multi Agent-based Simulation}

Agent-based modeling of complex systems involves representing each of the constituent participants of the system as an autonomous agent. These agents are designed to act and interact with one another in ways intended to lead to their aggregate behaviour approximating the modeled complex system. Such modeling has been applied to the study of financial markets, with market participants represented as agents trading in an artificial simulated market. Different types of traders and their various strategies can be separately modeled as agents and allowed to interact within the confines of the simulated market with the resulting dynamics intended to accurately reproduce a typical trading period in the market. Indeed certain stylized empirical facts and well known statistical regularities of financial markets such as the heavy-tailed distribution of asset returns, and volatility clustering \cite{cont2001empirical}, have been shown to emerge in such an agent-based simulation setting \cite{preis2006multi}. See also, \cite{paddrik2012agent} where agent-based modeling is used as a test-bed to test hypothesized causes of the 2010 `Flash Crash' phenomenon. A rich source of references on agent-based modeling techniques applied to the study of market dynamics in agent-based setting is \cite{lebaron2006agent}. \cite{wang2017spoofing} demonstrates an agent-based simulation of a market made up of 
what they refer to as \emph{Heuristic Belief Learning} agents. They show that such agents can be used to manipulate prices in such a market through spoofing, demonstrating the efficacy of using these agent-based approaches to understand such behaviours in real markets.


\subsection{Zero Intelligence Agents}

In our studies discussed below, we populate our IABS with hundreds of  simple agents, referred to as ``Zero Intelligence'' or ZI agents.
The term \emph{zero intelligence} agent was coined by \cite{gode1993allocative} to describe a family of automated market participants that submit random bid and ask orders. In this seminal work, two types of agents were considered: ZI-U (unconstrained) agents which place orders entirely at random within fixed extents, and ZI-C (constrained) agents which are prohibited from placing orders that result in an immediate loss. These ZI agents were initially used to demonstrate that the allocative efficiency of a market arises from its structure and not the particular strategy or intelligence of its participants, i.e., individual strategies are subsumed by the market as whole.

Subsequent use of ZI agents focused on the separation of market structure from participant strategy to allow isolated analysis of structural components.  For example, Bollerslev and Domowitz \cite{bollerslev_1993} use a set of ZI agents to analyze the structural impact of restricting the maximum depth of an order book.  Over time, the ``pure random'' aspect of ZI agents was relaxed to produce, for example, ``near zero intelligence'' agents which use recent mean transaction prices to study asset price bubbles and crashes \cite{duffy2006} and ``zero intelligence plus'' agents which maintain a value belief based on the recent order stream to improve market convergence under certain conditions \cite{cliff1997}.

Modern market simulations often use some form of ZI agent as a ``background'' agent to produce a reasonable baseline market microstructure into which experimental agents can be injected.  For example, Wang and Wellman's investigation of spoofing agents \cite{wang2017spoofing} uses a modified ZI agent with a Bayesian fundamental value belief based on noisy observations of an oracular value series, a private valuation per agent per unit, and a ``strategic parameter'' $\eta$ (eta) that controls the agent's willingness to accept less than its desired surplus in exchange for immediate, guaranteed execution.

For further information on the history of zero intelligence agents in agent-based computational economics, see the excellent review by Dan Ladley \cite{ladley_2012}.

\subsection{The ABIDES Interactive Agent-Based Simulator}

In this section, we refer to the ABIDES simulation framework \cite{simulator} with an obfuscated name and citation for the sake of peer review anonymity.  After peer review, references to ABIDES will be replaced with the proper framework name and the real citation will be included.

Both our backtesting and interactive simulation experiments utilize the open-source ABIDES framework, which provides an agent-based interactive discrete event simulator to support such investigations.  The ABIDES simulation platform provides support for continuous double-auction trading with nanosecond resolution, the ability to simulate specific dates in history either as a pure replay backtesting \emph{or} with gated agent access to noisy historical data, simulation of variable electronic network latency and agent computation delays, and a requirement that all agents intercommunicate solely through a standardized message protocol similar to that used by NASDAQ.  It also provides a hierarchy of heritable trading agent and exchange classes that facilitate rapid deployment of new experimental ideas.

We leverage ABIDES's ability to take the same historical market data, exchange agents, and trading agents and easily perform a head-to-head comparison of non-interactive backtesting results against interactive agent-based results.  For the current set of experiments, we compare historical backtesting to a population of zero intelligence (ZI) agents that make noisy observations of a stochastic mean-reverting process as the fundamental value of a single equity.  The historical backtest operates at real-world time scale and the agents based on the mean-reverting process operate in discrete time units.

\section{Approach}

\subsection{Backtesting in Agent-based Simulation}

One contribution of this paper is that we show how one can support sophisticated backtesting within an agent-based simulation using market replay.  In particular, we implement backtesting using three agents: An \emph{exchange agent} representing the exchange which keeps the order book (e.g., Nasdaq or NYSE), a \emph{ market replay agent} that provides liquidity by replaying historical orders and an \emph{experimental agent} representing the trading strategy to be evaluated.
As an example application, we investigate the impact of placing buy and sell market orders at different times and sizes to evaluate the short and long-term impacts on the mid price. 

The market replay agent uses real historical intra-day order data to replicate the evolution of the order book for a particular day. For our experiments, we used an order stream, implemented as a message stream,
%
%
containing a record of orders placed at different times. Each order is characterized by the time it was placed, the direction, size, price and order type (submissions, cancellations, partial cancellations, visible and hidden executions).  The data was provided by LOBSTER~\cite{LOBSTER}.


%
%

\begin{figure}[!htb]
 \centering
  \includegraphics[width=0.5\textwidth]{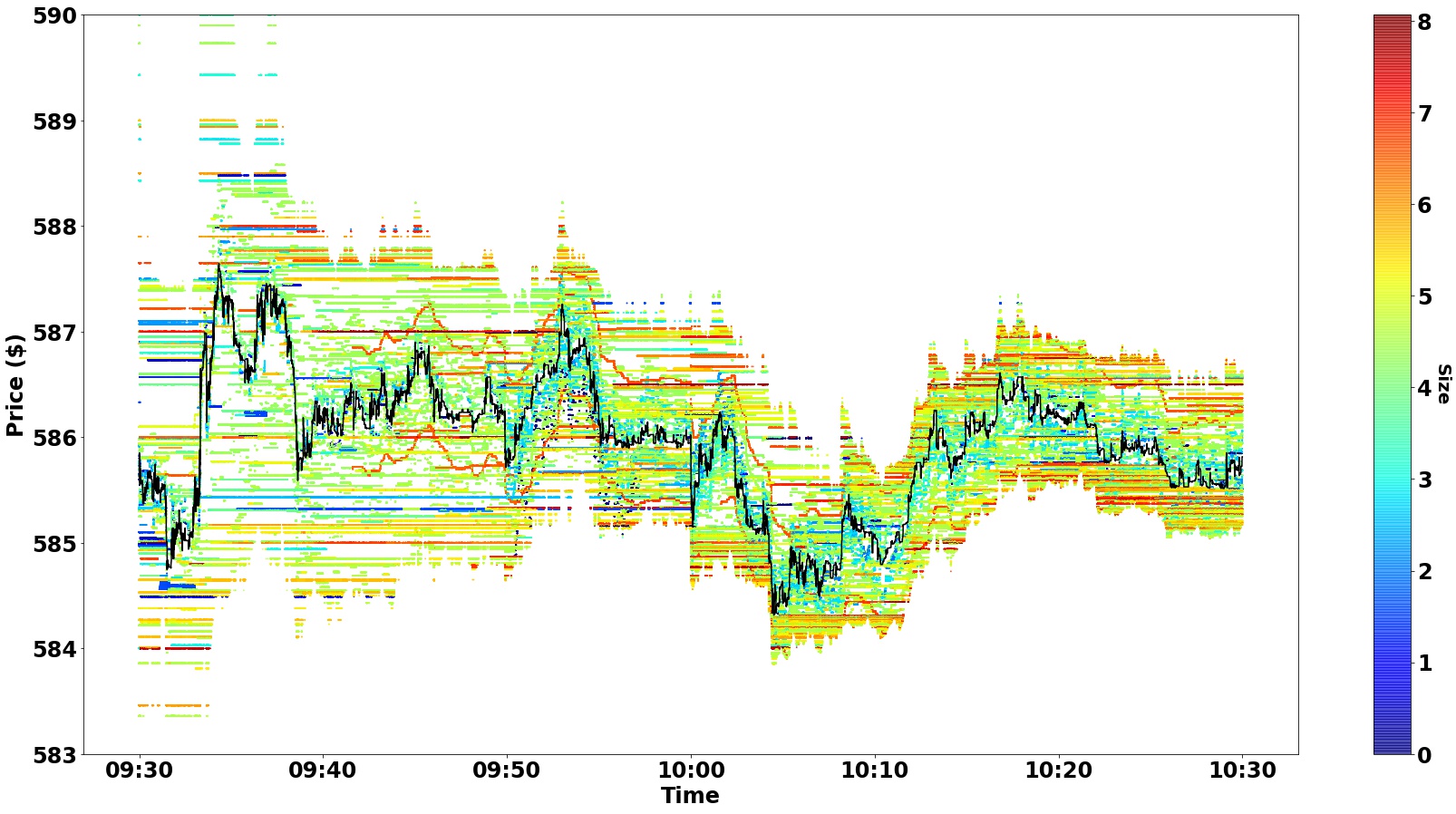}
 \caption{Price-level volume plot. Black line represents the mid price, Each point is the price at different price levels with the colour scheme indicating the size (log scale) present at each level}
 \label{fig:price_level_volumne}
 \end{figure}

The messages file contained 86,615 events representing all the different order types submitted between 09:30 and 10:30 with 41,844 unique order IDs. In terms of new limit order submissions, there were 22,222 sell limit orders and 19,333 buy limit orders. The average time between the 41,554 new limit order arrivals was 866 ms and the average time between the 38,791 cancellation and partial cancellation order arrivals was 928 ms. As part of our analysis, we visualize the evolution of the order book for all price levels and plot the price-level volume chart in figure \ref{fig:price_level_volumne}.

At the first time stamp available after the market opens, the historical order book file is referenced by the market replay agent to generate a list of new limit orders necessary to replicate the opening order book. This is needed as the orders stream does not contain the orders that led to the construction of the first order book snapshot. The submitted orders are handled by the exchange agent which uses an order book implementation and matching engine to update the order book based on the submitted and matched orders. After market open, the orders stream are processed sequentially and the orders are submitted to the exchange agent as simulated time reaches the time stamp associated with each historical order. Given that the orders stream contains visible and hidden execution messages, these had to be accounted for by either cancelling or partially cancelling the corresponding unfilled orders in the order book. 

The experimental agent is configured to participate in the simulation in a manner similar to the market replay agent, with the orders submitted dependant on the experiment carried out.

\begin{figure*}[t]
    \centering
    \subfloat[Experimental agent places \textbf{buy} order]{
        \includegraphics[width=0.5\textwidth]{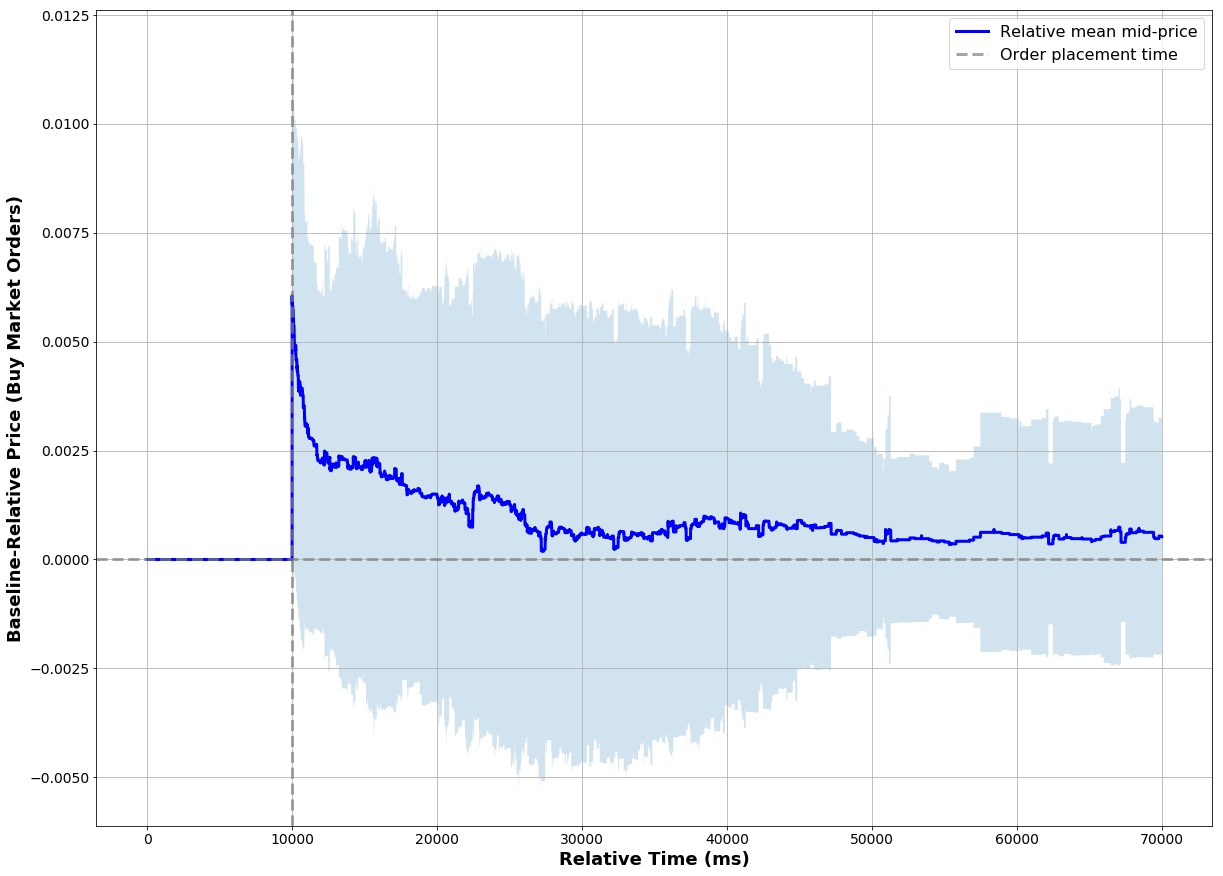}
    }
    \subfloat[Experimental agent places \textbf{sell} order]{
        \includegraphics[width=0.5\textwidth]{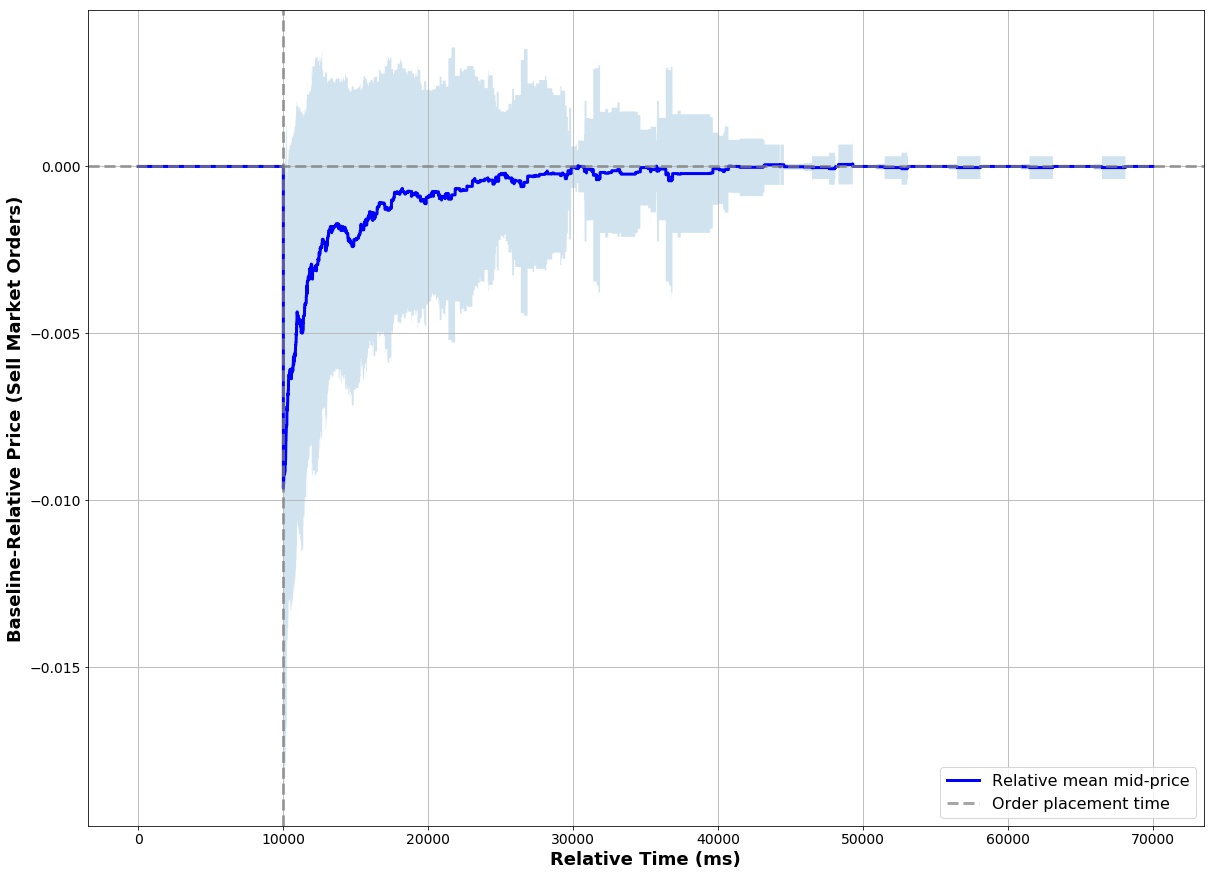}
    }
    \caption{Observed impact on the mid price by the experimental agent placing market orders at twice the best bid or ask size}
    \label{fig:backtest_impact_one_size}
\end{figure*}

\subsection{Event Studies}

As part of our methodology for evaluating experimental strategies, we leverage \emph{event studies}.
Event studies are so called because of their utility in analyzing the effect of a specific temporally-located event or class of events on a series of measures, such as the price quotes of an equity security.  There is a long history of event studies in economics and finance as related in Craig MacKinlay's excellent 1997 survey \cite{mackinlay1997}, which traces their use to at least 1933.

The general procedure to conduct an event study is as follows, with examples given at each step.  First, obtain a list of event times either exogenously or endogenously: for example identifying when news sources broke stories about executive indictments, or by computing the times at which an equity price series fell below some periodic moving average.  Second, ``cut out'' periods of the measure series in a window from shortly before to some time after each occurrence of the event: for example one day before to ten days after news breaks of corporate wrongdoing.  Third, align those subseries at the time of each event; for example placing each event at relative time zero, with negative X axis indicating times prior to the event.  Fourth, normalize each subseries to be relative to a benchmark level at the time of the event; for example dividing each equity price subseries by the price at the time of event, so all prices at ``time zero'' are exactly 1 and deviations from that price indicate cumulative returns from the event time.  Finally, combine the subseries event examples using an appropriate statistical or visual method to obtain some aggregated sense of the effect of the event on the measure series; for example, compute and visualize the mean and standard deviation of the set of normalized equity price subseries to understand the likely impact of a new occurrence of the event.

The event study is one of our primary tools to study the effect of the experimental ``impact trader'' in our current analysis.  We define the entry of the impact trader into the market to be the event and analyze across many different trials (times, dates, trade sizes) the ``typical'' effect of a single large trade on near-future price quotes for the same equity.



\begin{figure*}[t]
    \centering
    \subfloat[Experimental agent places \textbf{buy} order]{
        \includegraphics[width=0.5\textwidth]{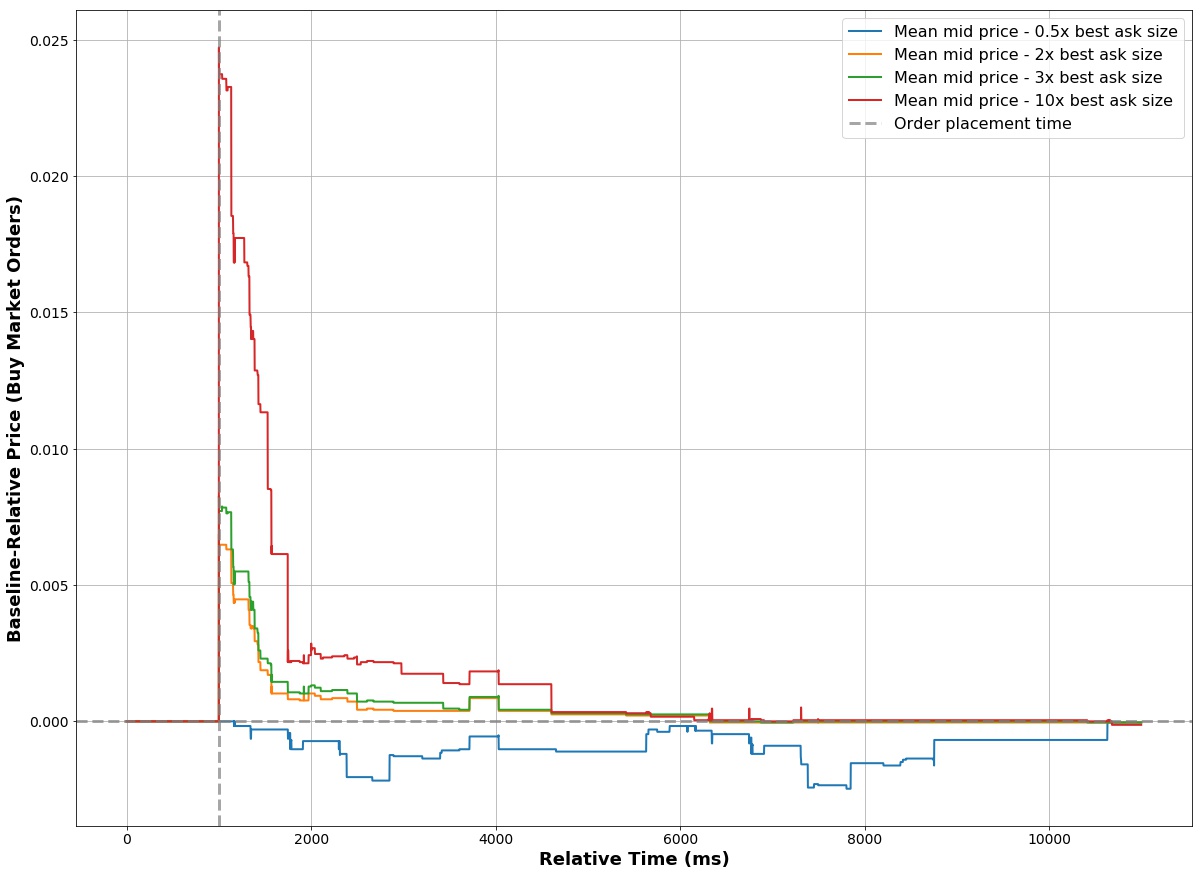}
    }
    \subfloat[Experimental agent places \textbf{sell} order]{
        \includegraphics[width=0.5\textwidth]{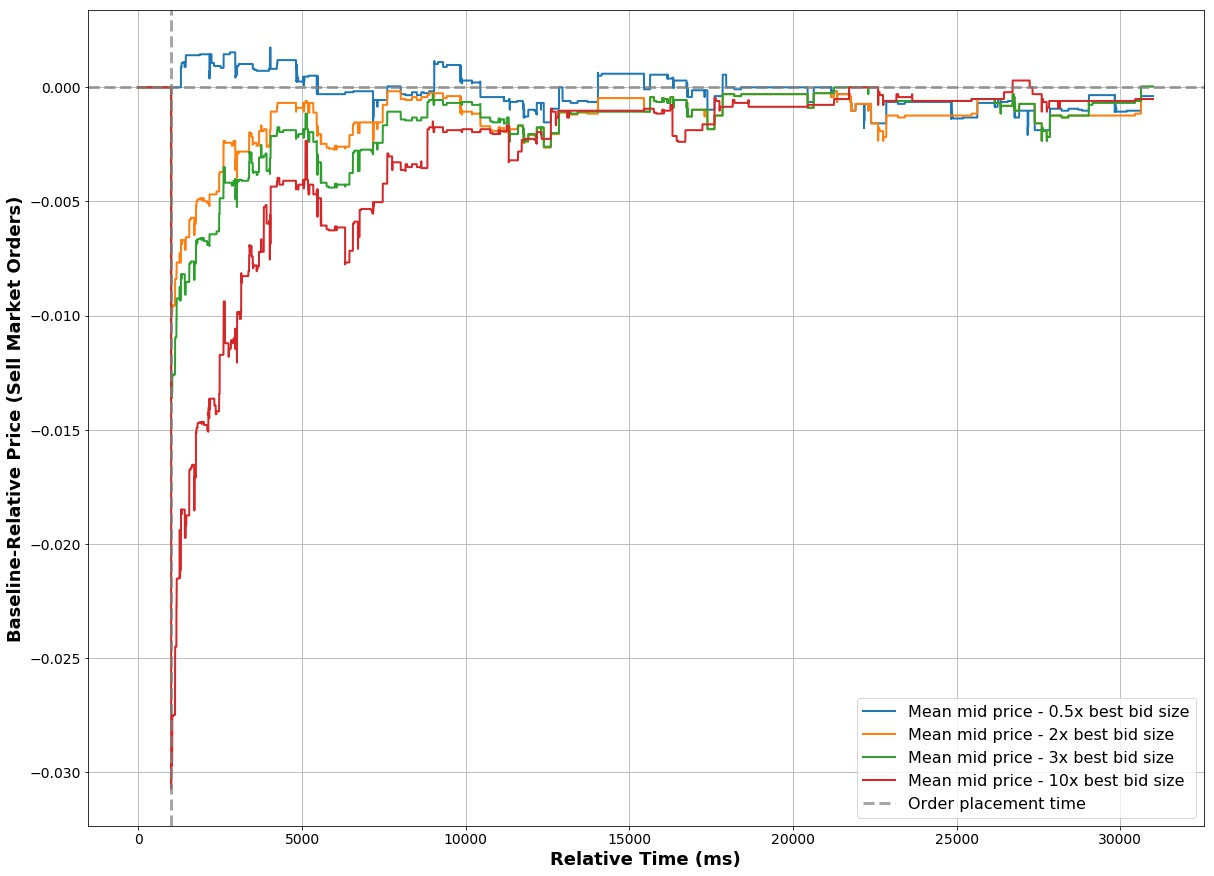}
    }
    \caption{Observed impact on the mid price by the experimental agent placing market orders at 50\%, 200\%, 300\% and 1000\% of the best bid or ask size}
    \label{fig:backtest_impact_mult_size}
\end{figure*}

\section{Backtester Market Impact Experiments}

The market impact studies take the form of a single experimental agent placing orders against the market replay agent providing liquidity during the simulation. In particular, we carry out two experiments: In the first experiment we evaluate the impact of a large market buy order by having the experimental agent place a buy order sized at 200\% of the best ask size. The bid/ask midpoint is recorded before and after the order and used in the analysis.  The experiment is repeated 100 times at different times of day, with the resulting data combined into a single event study.  We repeated the same process for sell orders.

The time range chosen for this experiment was between 09:45:00 and 09:53:15 with orders placed every 5 seconds providing 200 simulations in total (100 buys and 100 sells). Each simulation is run separately, with the experimental agent placing the orders within the time range chosen. This is to allow for assessing the impact on the mid price without any impact caused by other order placements. 
In subsequent experiments, we follow a similar setup. However, we vary the order sizes placed by the experimental agent. The time range chosen for this experiment was between 09:45:00 and 09:46:35 with orders placed every 5 seconds providing 40 simulations for each order size and 160 simulations in total for the order sizes chosen. The sizes are 50\%, 200\%, 300\% and 1000\% of the best ask or bid sizes.

We analyze the market impact by considering the mid price movement directly after the experimental agent order placement. We compare that against the baseline associated with the market replay agent placing orders without the presence of an experimental agent. In order to evaluate the different price impacts, we sample the price midpoints every 1 ms and normalize by the mid price associated with the market replay agent only. We then take the mean and standard deviation of the different simulations to produce figure \ref{fig:backtest_impact_one_size} showing the results of the first and second experiments and figure \ref{fig:backtest_impact_mult_size} showing the results of the subsequent experiments. For the first two experiments, we observe an immediate spike in the mid price followed by a decay converging towards the original mid price observed without the presence of an experimental agent.

%

In figure \ref{fig:backtest_impact_one_size}(a), we show the relative mid price 10 seconds and 60 seconds before and after the 100 market buy order placement times respectively. We also plot the standard deviation as a band around the mean to gain a good understanding of the different aggregated simulations. For the time range chosen, we observe an increase in the mid price in the form of a spike in the relative mean mid prices at the order placement times. This is expected given the nature of the market orders and how they remove the resting limit orders at the best ask levels. The relative mean mid price does not converge in our plots due to the time range chosen. However, when we look beyond 60 seconds after the orders placement, we observe that the mean mid price eventually converges. Similarly, in figure \ref{fig:backtest_impact_one_size}(b), we show the relative mid price 10 seconds and 60 seconds before and after 100 the market sell order placement times respectively. We observe a similar profile with the mid price decreasing and then converging over time. However, the rate of convergence is higher in the case of sell orders compared to buy orders. Another observation is around the standard deviation around the relative mean mid price whereby that of the sell market orders is narrower than that of the buy market orders. Finally, we note the presences of periodic relative mean mid price movements in the shape of pulses occurring after the convergence time stamp.

In figure \ref{fig:backtest_impact_mult_size}(a), we show the relative mid price 1 second and 10 seconds before and after the 80 market buy order placement times respectively. Clearly, the profiles associated with orders of higher sizes display higher increased mid prices and relatively longer time to converge. Interestingly, the mean-mid price associated with the experimental agent placing orders at 50\% of the best ask size results in a slight decrease in the relative mid price which eventually converges. Similarly, in figure \ref{fig:backtest_impact_mult_size}(b), we show the relative mid price 1 second and 10 seconds before and after the 80 market sell order placement times respectively. Again, the profiles associated with orders of higher sizes display lower decreased mid prices and relatively longer time to converge. We also note a similar pattern to that observed in figure \ref{fig:backtest_impact_mult_size}(a) whereby the mean-mid price associated with the experimental agent placing orders at 50\% of the best bid size results in a slight increase in the relative mid price.

\begin{figure*}[t]
    \centering
    \subfloat[Experimental agent places \textbf{buy} order]{
        \includegraphics[width=0.5\textwidth]{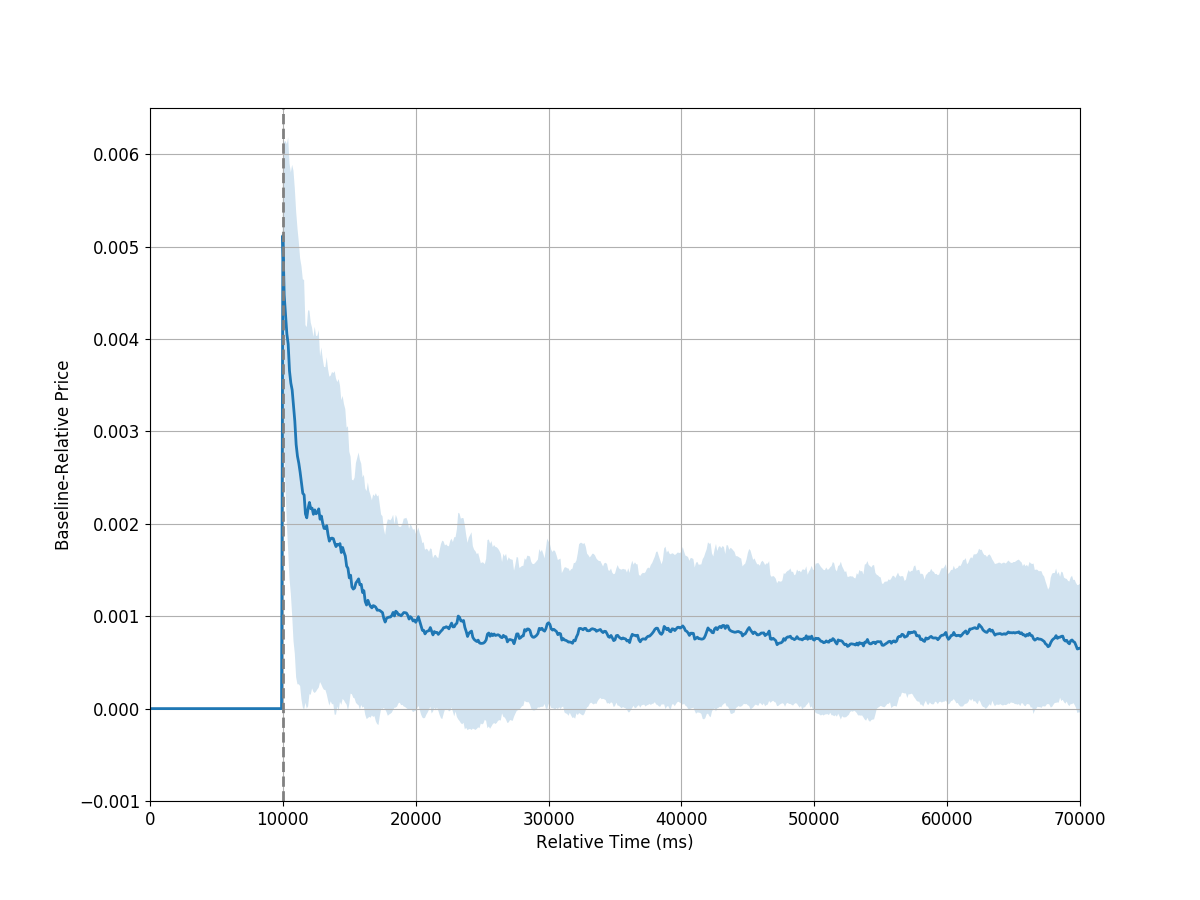}
    }
    \subfloat[Experimental agent places \textbf{sell} order]{
        \includegraphics[width=0.5\textwidth]{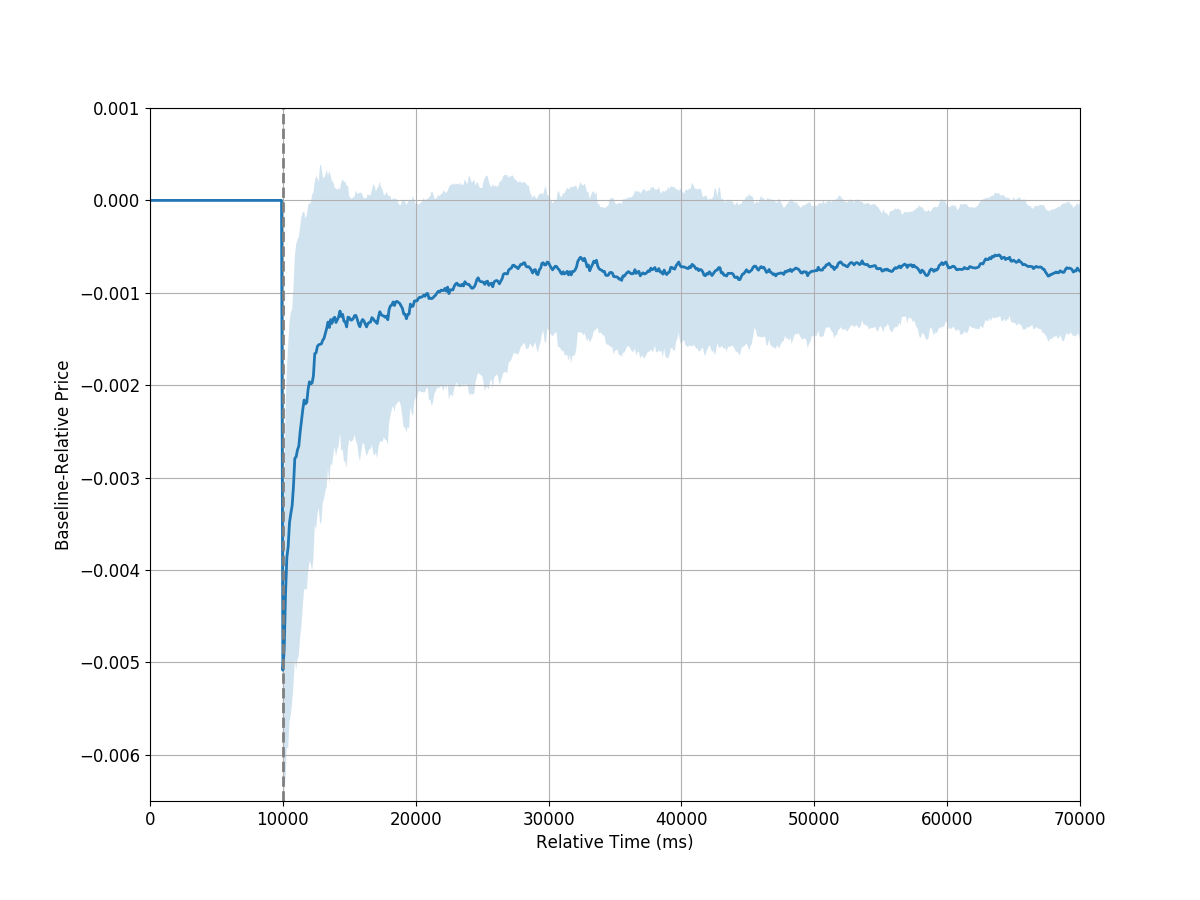}
    }
    \caption{Observed impact on the mid price by the experimental agent placing market orders with $greed=1.0$}
    \label{fig:iabs_one_size}
\end{figure*}

\begin{figure*}[t]
    \centering
    \subfloat[Experimental agent places \textbf{buy} order]{
        \includegraphics[width=0.5\textwidth]{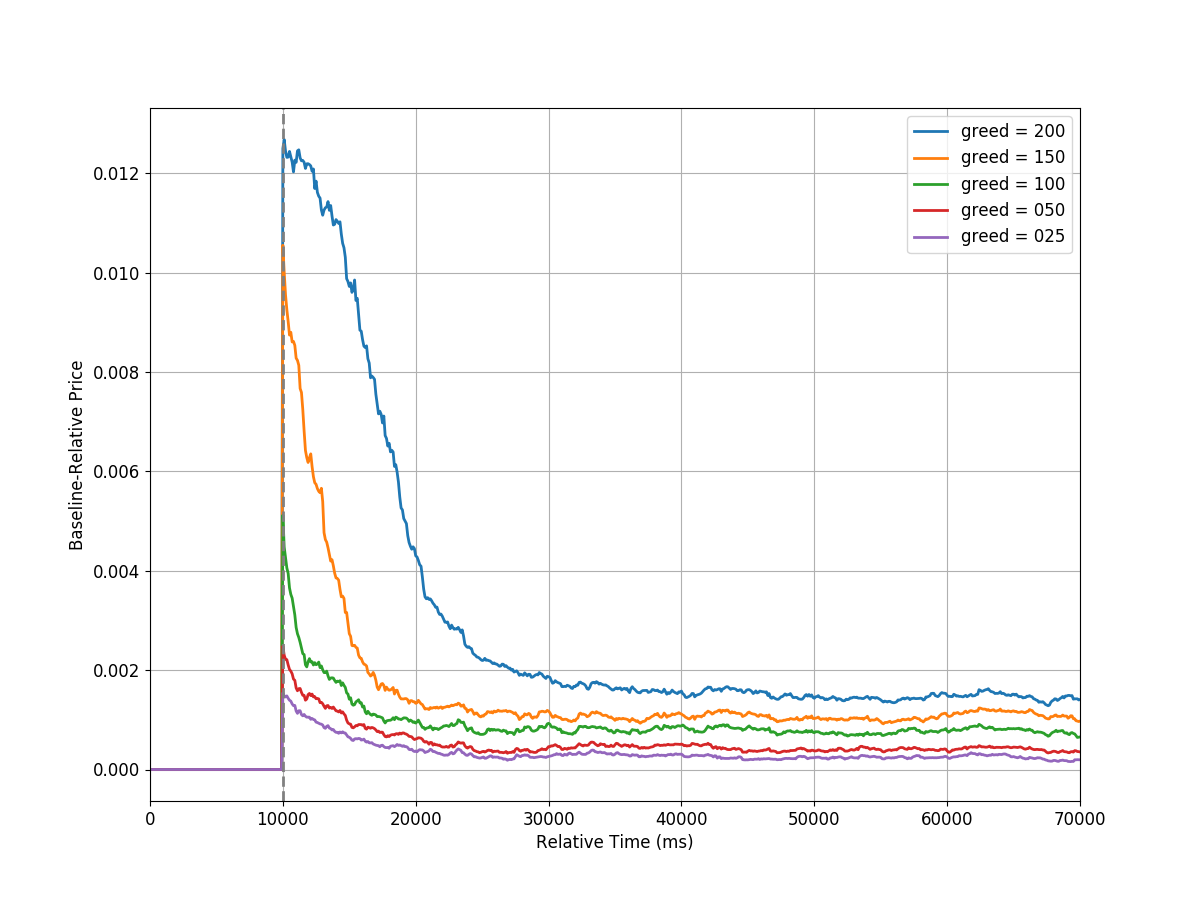}
    }
    \subfloat[Experimental agent places \textbf{sell} order]{
        \includegraphics[width=0.5\textwidth]{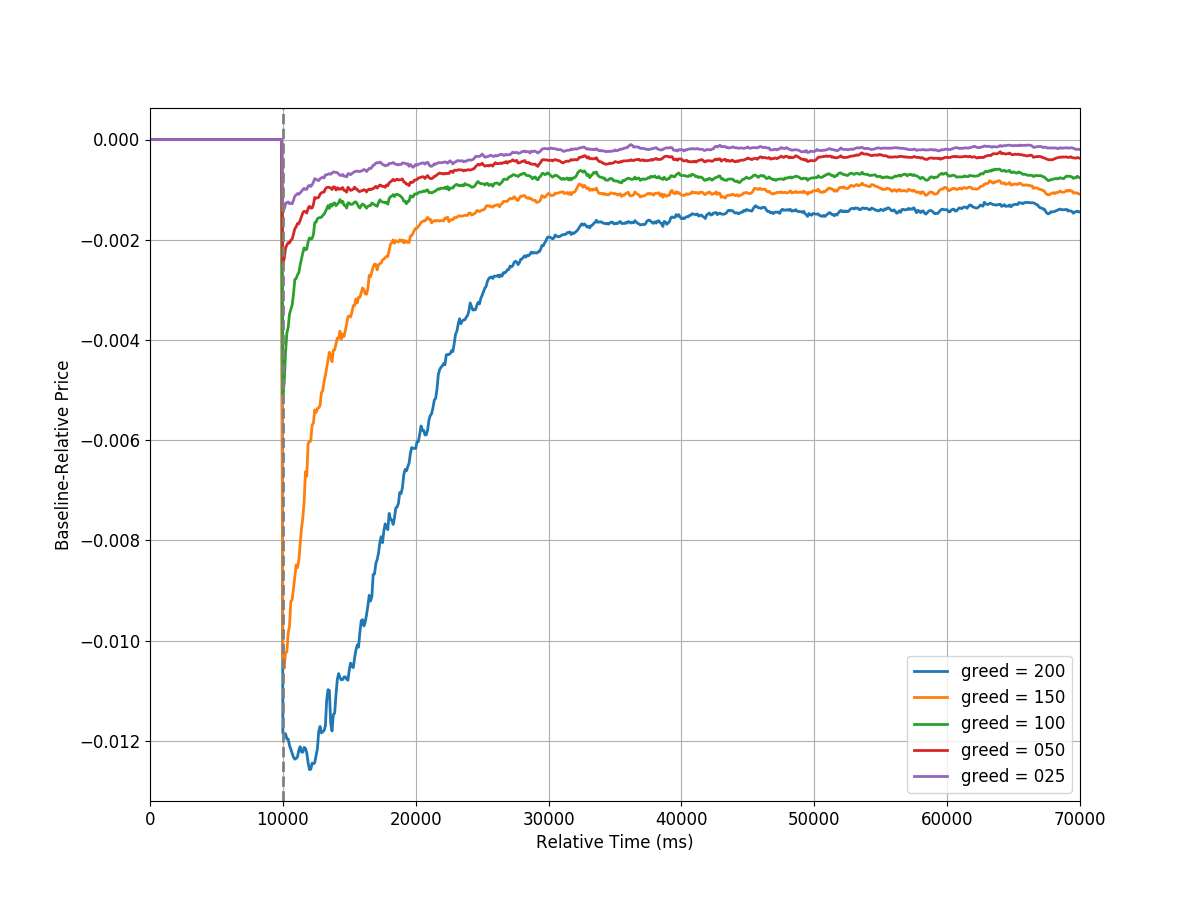}
    }
    \caption{Observed impact on the mid price by the experimental agent placing market orders with varying $greed$}
    \label{fig:iabs_mult_size}
\end{figure*}

\section{IABS Market Impact Experiments}

For this series of experiments, following the approach taken by others using evaluations with ZI agents, the interactive market simulation was configured to proceed in discrete time units with 1,000 total time units per simulation. (In future work we intend to adapt ZI agents to continuous time markets.) Each agent arrived at the market according to a Poisson process, thus not every agent could act within every time unit.  The exchange processed messages at each time unit, with messages in the same time unit handled in arbitrary order.  A single equity was available to trade.  Its fundamental value sequence, which we think of as the unobservable true consensus value of the equity, was taken to be a stochastic mean-reverting process.  Participating agents received noisy observations of this fundamental value at each market arrival.


The experimental agent in this study is the ``impact agent'' which places a large, single order into the market at a particular time, with size determined by available liquidity and a ``greed'' parameter.  The background agent population engaged to study market impact consists of 100 zero intelligence (ZI) agents which randomly decide to buy or sell at market arrival, then select a limit price chosen to produce a specific requested surplus to the agent if successfully transacted.  The requested surplus extents are a strategic parameter that varies among the agent population, as is the strategic threshold $\eta$ (eta) as per Wang and Wellman's construction \cite{wang2017spoofing}.  Our ZI agents maintain a Bayesian belief about the current fundamental value and current error, and use this to estimate the final fundamental value as their anchor point for considering expected surplus.

The simulation is carefully constructed such that given the same global random seed, every background agent will receive the same random number sequence whether or not the experimental agent is present.  Deviations in background agent behavior should therefore arise from experimentally-altered market conditions and not simple random perturbation of their decision processes.  Given this feature, each instance of simulation was conducted twice in an experimental-control pair using the same global random seed, once with the experimental impact agent trading as designed, and once with the impact agent present but not actually placing its ``impactful'' order.  The difference, at each discrete time unit, in the bid-ask midpoint between the experimental trial and the control is considered the ``impact'' of the experimental trader at that time unit.


In the simulation experiment, the experimental agent places the ``impact order'' at time unit $t=200$ in a simulation of 1,000 time units duration.  Each background ZI agent first arrives at the market in the range $t\in[0,100]$ and then following a Poisson process.  Thus the equity order book will be well populated by the time of the impact event.  The impact agent queries the exchange for liquidity data in the form of price levels and aggregated volumes near the spread.  It uses a strategic ``greed'' parameter to determine what size order to place relative to the available liquidity.  In the visualized experiment, the impact agent queries liquidity within 1\% of the inside bid (if selling) or ask (if buying) and with $greed=1.0$ places an order to capture all of it.

The current simulation configuration operates in arbitrary discrete time units.  For the sake of comparison with the backtest, we capture impact from 100 units before each event ($T-100$) to 600 units after each event ($T+600$), and we consider each arbitrary time unit to represent 100 milliseconds.

We conducted 100 experimental-control trials (200 distinct simulations) as described above and aggregated the results in the form of an event study examining the effect of the impact order on subsequent bid-ask quote midpoints.  We compute the aggregated effect of the impact order as the percent change between each trial's control price series and experimental price series and visually present the mean and standard deviation of those changes in Figure \ref{fig:iabs_one_size}.  We conducted a similar set of trials while varying the impact agent's greed parameter and present the mean observed impact by greed in Figure \ref{fig:iabs_mult_size}.

We note that in the interactive agent based simulation, the experimental mean price does \emph{not} return to the baseline price, for which we offer three explanations.

First, even if our background ZI traders did not alter their subsequent actions based on the impact trade (as the backtest obviously does not), the makeup of the order book is now different with regions near the spread substantially thinned out and unavailable for trading.  This means orders that may have immediately executed (against volume ``stolen'' by the impact trader) will now enter the order book instead.  However, as the ZI agents cancel all outstanding orders upon each market arrival anyway, this would explain only very short term effects until all ZI agents had again arrived at the market.

Second, the ZI agents offered limit prices are computed in part on a private value vector representing each agent's unique valuation of each unit of stock (with diminishing returns).  Because the impact agent has executed trades that were not executed in the control, the ZI agents' holdings will have changed, causing different individual valuations for acquiring additional shares.

Third, even if the ZI agents did not select different limit prices to offer, the presence of their strategic parameter eta ($\eta$) can cause them to trade at the current simulated spread instead of their desired limit price, giving up some surplus for guaranteed, immediate execution.  Thus a change in the inside bid-ask quotes can alter their behavior.

As illustrated by the simulation experiments, the ability of market participant agents to react to the event allows the market to reach a new, different equilibrium that could not be realized in the backtest environment.

\section{Conclusion and Future Work}

We have shown how an existing Interactive Agent-based Simulation market environment can be adapted for use as a backtester.  This approach enables the same candidate/experimental trading strategy to be evaluated in a backtesting context as well as in an IABS context. To illustrate the process and to investigate differences in outcomes, we evaluated a very simple experimental trading agent, namely one that enters a single, large market order.  We use the event study methodology to evaluate and compare the price impact caused by this order.

We had not mentioned it before, but we should point out that  it is not feasible in a real market to conduct experiments that definitively evaluate the impact of orders on the market: We can never know how the market would have proceeded without the experimental order being present.  As such, it is difficult to know for certain which of the two methods we explored provides a more realistic background for evaluation.  However, the consensus in the literature indicates that substantial orders cause an initial ``shock'' impact with a gradual decay.  Some models suggest that in some cases the effect is permanent, and the price will never revert to the baseline. \cite{alfonsi2010optimal}.

We observe in our experiments that in the backtesting environment the price trends rather quickly back to the baseline price, eventually reaching that price and remaining there for the {\bf sell} experiments. In the backtesting {\bf buy} experiments, the price trends towards, but does not stabilize at the baseline price. In the IABS experiments, however, we see the price stabilize at a new level in each set of experiments, suggesting that the impact of the order is longer lasting or even permanent. This corresponds more closely with existing impact models and it makes sense because the ZI agents acting as background traders do maintain state that can be permanently affected by a market event and modify their future behavior.

It is not clear at this stage that one approach is certainly better than the other, but this paper shows that both evaluation techniques can be pursued in the context of a single IABS framework. We hope to continue our work on this testbed. Our results suggest a number of questions to explore soon. Among other things, we hope to extend the realism and diversity of background trading agents for IABS.  A first step in that direction will be to break away from our implementation of ZI traders that requires a fixed time step approach to support continuous time markets.


\section*{Acknowledgements}

This material is based on research supported in part by the National Science Foundation under Grant no. 1741026, and by a JPMorgan AI Research Fellowship.

\section*{J.P. Morgan Disclaimer}

Opinions and estimates constitute our judgement as of the date of this paper, are for informational purposes only and are subject to change without notice. This paper is not the product of J.P. Morgan’s Artificial Intelligence Research Department and therefore, has not been prepared in accordance with legal requirements to promote the independence of market research, including but not limited to, the prohibition on the dealing ahead of the dissemination of investment research. This paper is not intended as market research, a recommendation, advice, offer or solicitation for the purchase or sale of any financial product or service, or to be used in any way for evaluating the merits of participating in any transaction. It is not a market research report and is not intended as such. Past performance is not indicative of future results. Please consult your own advisors regarding legal, tax, accounting or any other aspects including suitability implications for your particular circumstances. J.P. Morgan disclaims any responsibility or liability whatsoever for the quality, accuracy or completeness of the information herein, and for any reliance on, or use of this material in any way.

\newpage
\bibliography{bib}
\bibliographystyle{icml2019}

\end{document}